\documentclass[runningheads]{llncs}
\usepackage{amsmath, amssymb}
\usepackage{booktabs}
\usepackage[T1]{fontenc}
\usepackage[utf8]{inputenc}
\usepackage[hyphens]{url}
\usepackage{caption}

\usepackage{cite}

\usepackage{makecell} 
\usepackage{array}    

\usepackage{graphicx}
\usepackage{svg}
\usepackage{url}
\usepackage{enumitem}
\usepackage[ruled,vlined,linesnumbered]{algorithm2e}

\usepackage{float}  
\usepackage{microtype}  
\begin{document}

\title{DermETAS-SNA LLM: A Dermatology Focused Evolutionary Transformer Architecture Search with StackNet Augmented LLM Assistant}

\titlerunning{DermETAS-SNA LLM Assistant}
%
%
\author{Nitya Phani Santosh Oruganty\inst{1}*\orcidID{0009-0009-0491-8162} \and Keerthi Vemula Murali\inst{2}*\orcidID{0009-0009-3868-4376} \and Chun-Kit Ngan\inst{1}\orcidID{0000-0003-2151-0459} \and Paulo Bandeira Pinho \inst{3}
}
\authorrunning{Santosh, Keerthi, et al.}
%
\institute{
Data Science Program, Worcester Polytechnic Institute, Massachusetts, USA 
\and
Computer Science Department, Worcester Polytechnic Institute, Massachusetts, USA 
\and
PASE Advisory Group, New Jersey, USA
}


%
\def\thefootnote{*}\footnotetext{These authors contributed equally to this work}\def\thefootnote{\arabic{footnote}}

\maketitle 

\begin{abstract}
This work introduces the DermETAS-SNA LLM Assistant that integrates Dermatology-focused Evolutionary Transformer Architecture Search (ETAS) with StackNet Augmented LLM. The assistant dynamically learns skin-disease classifiers and provides medically informed descriptions to facilitate clinician-patient interpretation. Contributions include: (1) Developed an ETAS framework on the SKINCON dataset to optimize a Vision Transformer (ViT) tailored for dermatological feature representation and then fine-tuned binary classifiers for each of the 23 skin disease categories in the DermNet dataset to enhance classification performance; (2) Designed a StackNet architecture that integrates multiple fine-tuned binary ViT classifiers to enhance predictive robustness and mitigate class imbalance issues; (3) Implemented a RAG pipeline, termed Diagnostic Explanation and Retrieval Model for Dermatology, which harnesses the capabilities of the Google Gemini 2.5 Pro LLM architecture to generate personalized, contextually informed diagnostic descriptions and explanations for patients, leveraging a repository of verified dermatological materials; (4) Performed extensive experimental evaluations on 23 skin disease categories to demonstrate substantial performance increase, achieving an overall F1-score of 56.30\% that notably surpasses SkinGPT-4 (48.51\%) by a considerable margin, representing a  performance increase of 16.06\%; (5) Conducted a domain-expert evaluation, with eight licensed medical doctors, of the clinical responses generated by our AI assistant for seven dermatological conditions. Our results show a 92\% agreement rate with the assessments provided by our AI assistant, demonstrating superior performance compared to SkinGPT-4 with a substantial margin (48.20\%); and (6) Created a proof-of-concept prototype that fully integrates our DermETAS-SNA LLM into our AI assistant to demonstrate its practical feasibility for real-world clinical and educational applications.

\end{abstract}
\section{Introduction}
The skin, being the largest organ of the body, is crucial for regulating internal temperature, retaining body fluids, and acting as the first line of defense against harmful pathogens. Despite its significance, the incidence of skin and subcutaneous diseases, including warts, actinic keratosis, eczema, seborrheic keratoses, psoriasis, onychomycosis (nail fungus), alopecia (hair loss) and more, is increasing worldwide and has been officially recognized by the World Health Assembly \cite{who2025skin} as a major public health issue. Atopic dermatitis (AD) is among the top 15 nonfatal global diseases, with a higher prevalence in urban and industrialized areas \cite{Tian2023}. Warts, caused by HPV, affect 7–12\% worldwide, especially school-aged children \cite{ref3}. Actinic keratosis (AK), a precancerous lesion, impacts 14\% people worldwide, yet 85\% are unaware, its treatment market is projected to rise from USD 7.95B (2024) to USD 13.0B by 2032 \cite{actinic}. Seborrheic keratosis (SK), the most common benign skin tumor, affects almost all individuals over 60 years of age and 12\% of those aged 15-30 in the United States \cite{ref6}. Nail disorders such as onychomycosis represent 50\% of nail diseases, increasing with age \cite{ref7}. Skin cancer is a significant concern with more than 5.4M US non-melanoma cases yearly, and melanoma is projected to cause 8,430 deaths by 2025 \cite{ref_article1}.

As per the 2021 Global Burden of Disease Study \cite{who2025skin}, there were 4.69 billion new cases of skin diseases globally, ranking them among the top ten causes of disability. This growing burden is exacerbated by a widespread lack of understanding of skin diseases in various communities, which often leads to delayed diagnoses and insufficient treatment. The World Health Organization (WHO) \cite{who2025skin} highlights that many communities face a shortage of trained healthcare professionals and specialists, which further hampers the effective management of skin conditions. For instance, as of 2023, the United States had 12,120 active dermatologists, (3.66 per 100,000 people), an increase from 10,845 in 2016. Nevertheless, the issue of maldistribution has intensified, with 48.4\% of dermatologists concentrated in the 100 most densely populated ZIP codes, while only 0.9\% cater to the 100 least densely populated areas. Despite the growth of the workforce, the geographic disparity is increasing, indicating an urgent need for policies aimed at enhancing access to dermatological services in underserved communities \cite{Shah2024}.

To address the needs, the recent advances in Artificial Intelligence (AI), including computer vision, large language models (LLMs), and multimodal models, have resulted in a swift expansion of medical applications, particularly in dermatology. SkinGPT-4 \cite{skingpt2024} is an interactive dermatology diagnostic system based on multimodal LLMs that can deliver a practical solution to the global shortage of dermatologists and accessibility challenges by enabling diagnosis using a Vision Transformer (ViT) \cite{vit} aligned with Meta's LLaMA-2-13B Chat model \cite{llama2}. However, SkinGPT-4 exhibits certain constraints and limitations. First, it uses a single model for all skin disease categories, which may reduce diagnostic precision for rare or underrepresented conditions. Second, it lacks direct integration with authoritative dermatological knowledge repositories, like medical documents and clinical guidelines, limiting interpretability and traceability of its responses. As with other LLM-based systems, it risks generating hallucinated, overconfident, and clinically unsafe outputs without expert validation.

To overcome the existing limitations, we introduce our DermETAS-SNA LLM assistant, a unified system integrating Dermatology-focused Evolutionary Transformer Architecture Search with StackNet-Augmented LLM. Specifically, this work presents six technical contributions:

\begin{enumerate}
    \item We developed an \textit{Evolutionary Transformer Architecture Search} (ETAS) framework on the \textit{SKINCON} dataset \cite{skincon} to optimize a ViT architecture \cite{vit} tailored for dermatological feature representation. The optimized architecture is subsequently fine-tuned as a set of class-specific binary classifiers for each of the 23 disease categories in the \textit{DermNet} dataset \cite{dermnet}, thereby enhancing performance for rare and underrepresented classes.
    \item We designed \textit{StackNet}, a modular stacking ensemble framework that integrates multiple fine-tuned, class-specific binary ViT classifiers to enhance predictive robustness and mitigate class imbalance issues.
    \item We implemented a \textit{Retrieval-Augmented Generation} (RAG) \cite{rag} pipeline, termed DERM-RAG (Diagnostic Explanation and Retrieval Model for Dermatology), which harnesses the capabilities of the \textit{Google Gemini 2.5 Pro Preview-05-06} \cite{gemini25pro} LLM architecture. This pipeline generates personalized, contextually informed diagnostic descriptions and explanations for patients, leveraging a repository of verified dermatological materials.
    \item We performed extensive experimental evaluations on 23 disease categories within the DermNet dataset. Our AI assistant demonstrates strong performance, achieving an overall F1-score of 56.30\% that notably surpasses SkinGPT-4 (i.e., 48.51\%) by a substantial performance increase of 16.06\%.
    \item We conducted a domain-expert evaluation of the clinical responses generated by our AI assistant for seven dermatological conditions including Warts, Actinic Keratosis, Eczema, Seborrheic Keratoses, Psoriasis, Nail Fungus, and Hair Loss. The eight licensed medical professionals evaluated the quality and validity of the responses using a six-question survey instrument. Our results show a 92\% agreement rate with the assessments provided by our AI assistant, demonstrating superior performance compared to SkinGPT-4 with a substantial margin (i.e., 48.20\%). This outcome highlights the effectiveness of our AI assistant in providing accurate dermatological decision support.
    \item We created a proof-of-concept prototype that fully integrates our DermETAS-SNA LLM into our unified multimodal AI assistant to demonstrate its practical feasibility for real-world clinical and educational applications.

\end{enumerate}


The remainder of this paper is structured as follows. Section 2 provides a detailed overview of \textit{SkinGPT-4}. Section 3 describes the skin disease datasets, medical documents, and supporting resources used in our work and experiments. Section 4 introduces our proposed AI assistant. Section 5 presents the ETAS approach. Section 6 details the \textit{StackNet} architecture. Section 7 explains the DERM-RAG pipeline. Section 8 discusses our experimental evaluations and domain-expert assessments. Section 9 describes the proof-of-concept prototype implementation. Section 10 concludes the paper and outlines directions for potential future work to further enhance the proposed system.


\section{SkinGPT-4 System}


\begin{figure}[H]
  \centering
    \includegraphics[width=0.76\textwidth]{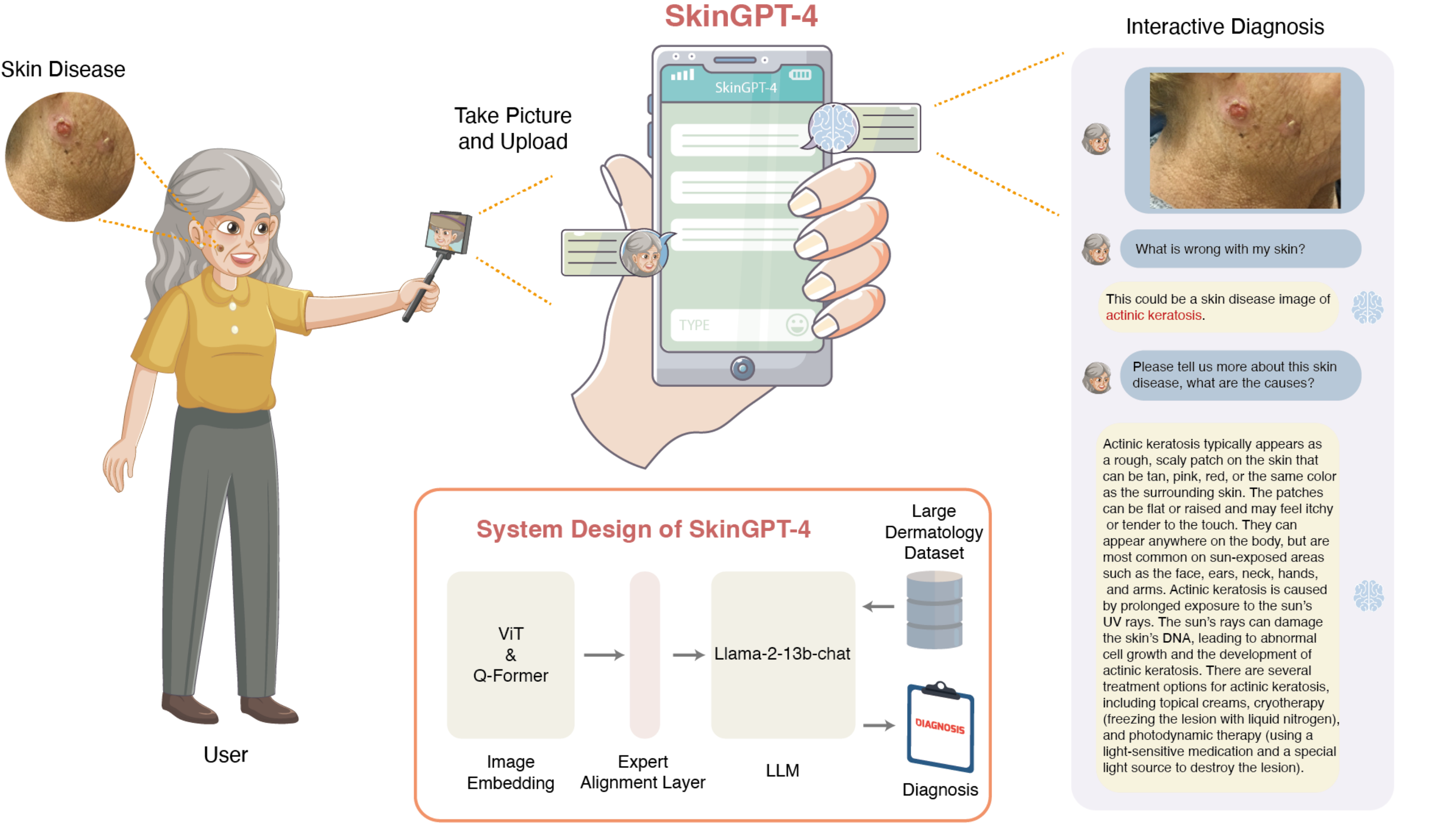}
    \captionof{figure}{SkinGPT-4 System Design and Interactive Diagnosis Example\cite{skingpt2024}}
    \label{fig:skingpt_architecture}
\end{figure}

As shown in Fig.~\ref{fig:skingpt_architecture}, SkinGPT-4 \cite{skingpt2024} is a multimodal LLM framework for dermatological diagnosis, aligning a pre-trained ViT with Meta's LLaMA-2-13B Chat model. It is trained in two steps: STEP 1 teaches the model to recognize and describe clinical features using 52,929 image-concept pairs, and STEP 2 focuses on disease-specific classification with 49,043 image-description pairs. Users can upload skin images, which are analyzed for visual features and classified into 15 disease categories derived from the original 23 DermNet classes, grouped by board-certified dermatologists. The model then provides interactive diagnostic explanations and treatment recommendations, supporting real-time self-diagnosis and doctor-patient communication.

SkinGPT-4 addresses major bottlenecks in dermatology, including long wait times, specialist shortages, and the need for explainable AI. Evaluated on 150 real-world cases reviewed by board-certified dermatologists, it achieved over 80\% diagnostic agreement and was considered informative, useful, and privacy-conscious. It supports local deployment, making it accessible for underserved regions. While not a replacement for physicians, SkinGPT-4 serves as a valuable triage tool and assistant, capable of 24/7 personalized diagnostics. 


SkinGPT-4 uses a single model to classify all 23 skin disease categories, simplifying deployment but potentially compromising precision, particularly for rare or underrepresented conditions. Diseases such as poison ivy, urticaria, atopic dermatitis, scabies, bacterial infections such as impetigo and cellulitis, and viral or STD-related conditions such as Herpes and HPV may suffer from a low representation in training data, leading to generalization errors, misclassification, and lower confidence in predictions. In addition, the model does not integrate external trusted dermatology literature or clinical guidelines, which limits the traceability and clinical reliability of its results. This absence of authoritative grounding increases the risk of hallucinated or overconfident diagnostic responses, which poses significant risks when the system is used without expert oversight.

\section{Skin Disease Datasets and Medical Corpus}
\subsection{SKINCON Dataset}

The SKINCON \cite{skincon} dataset is a densely annotated dermatological dataset designed to support fine-grained model interpretability and debugging. It consists of 3,230 images from the Fitzpatrick17k dataset and 656 images from the Diverse Dermatology Images (DDI) dataset, making a total of 3,886 images. Each image is labeled with 48 clinical concepts, including Vesicle, Papule, Macule, Plaque, Abscess, and more features related to pigmentation, texture, shape and lesion morphology. These annotations were developed by board-certified dermatologists using a standardized lexicon derived from \textit{Dermatology by Bolognia et al. (2017)}. This dataset enables training concept bottleneck models, probing representations, and generating concept-based explanations.


\subsection{DermNet Dataset}
The DermNet dataset \cite{dermnet} is a comprehensive collection of clinical dermatology images designed for multi-class classification tasks. It consists of 19,500 images meticulously categorized into 23 distinct skin disease categories, including Acne, Actinic Keratosis, Atopic Dermatitis, Cellulitis, Exanthems \& Drug Eruptions, Hair Loss, Herpes, Pigmentation Disorders, Lupus, Melanoma, Nail Fungus, Eczema, Bullous Disease, Poison Ivy, Psoriasis, Scabies, Seborrheic Keratoses, Systemic Disease, Tinea Ringworm, Urticaria Hives, Vascular Tumors, Vasculitis, and Warts. In this study, the DermNet dataset was used to train the StackNet classifier and to evaluate downstream diagnostic accuracy following feature transfer from SKINCON-trained encoders.

\subsection{Medical Corpus}
To supplement visual learning with textual expertise, we constructed a domain-specific corpus by extracting relevant information from five authoritative dermatology sources: (1) Habif’s Clinical Dermatology: A Color Guide to Diagnosis and Therapy \cite{habif2021},(2) Rook’s Textbook of Dermatology \cite{rook2017},(3) Textbook of Primary Care Dermatology\cite{primarycare2021},(4) Fitzpatrick Dermatology in General Medicine\cite{fitzpatrick2008}, and (5) The Electronic Textbook of Dermatology\cite{electronic}.This curated corpus served as the knowledge base in our DERM-RAG module to generate clinically grounded explanations for dermatology-related queries.

\section{DermETAS-SNA LLM Assistant}
\vspace{-2mm}

\begin{figure}[H]
\centering
  \includegraphics[width=0.85\textwidth]{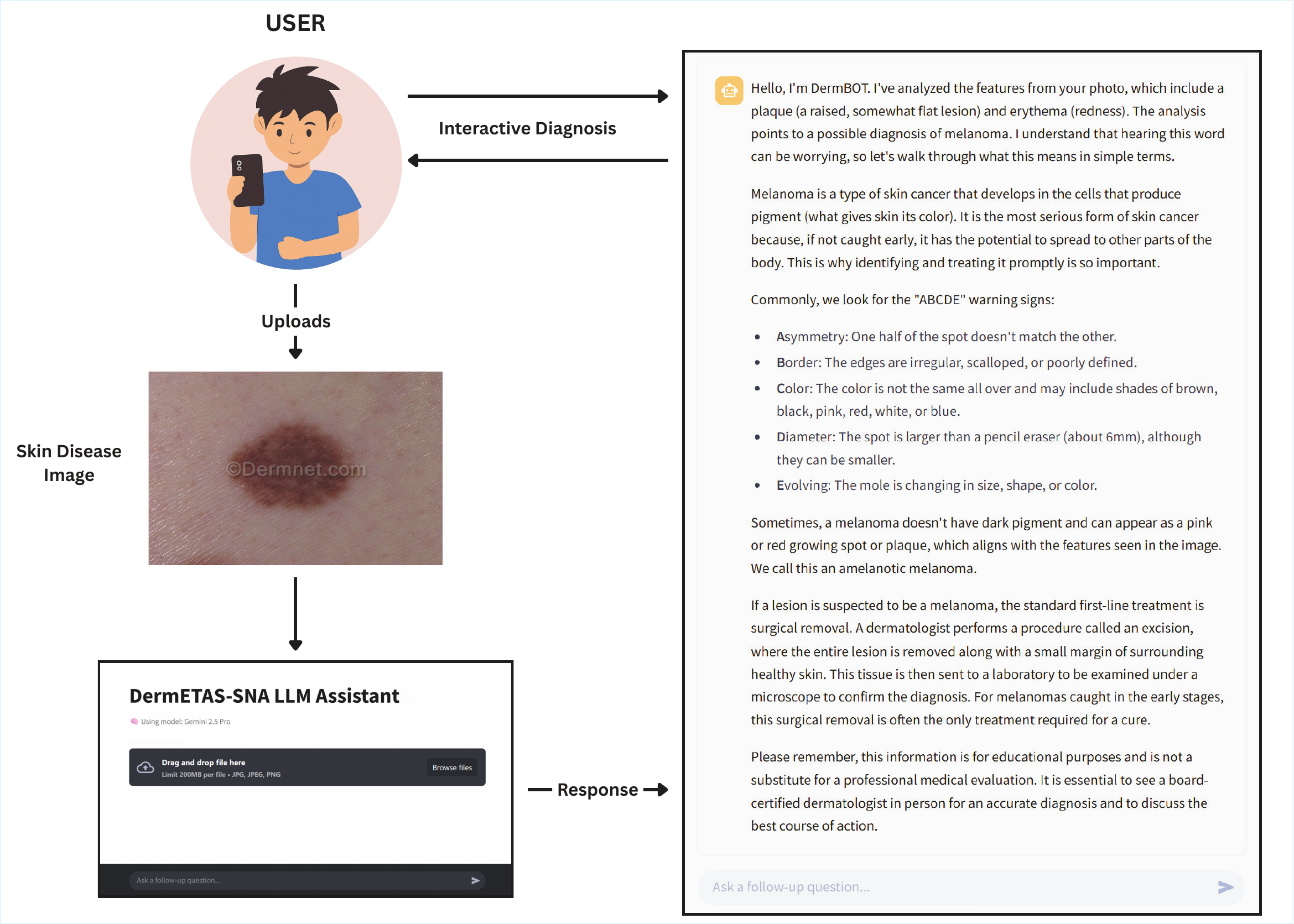}
    \captionof{figure}{Overview of the DermETAS-SNA LLM Assistant Interaction Flow}
  \label{fig:main}
  \end{figure}

Fig.\ref{fig:main} shows how users interact with our application. We present a modular and scalable framework designed to address key limitations of current dermatological AI systems, including architectural inefficiency, poor performance in underrepresented classes, limited explainability, and the absence of grounded diagnostic reasoning. Our approach comprises two tightly integrated phases: (1) Optimized vision-based diagnosis using customized Transformer architectures and (2) Generation of contextual explanations using DERM-RAG.

\begin{figure}[t]
\centering
  \includegraphics[width=0.70\textwidth]{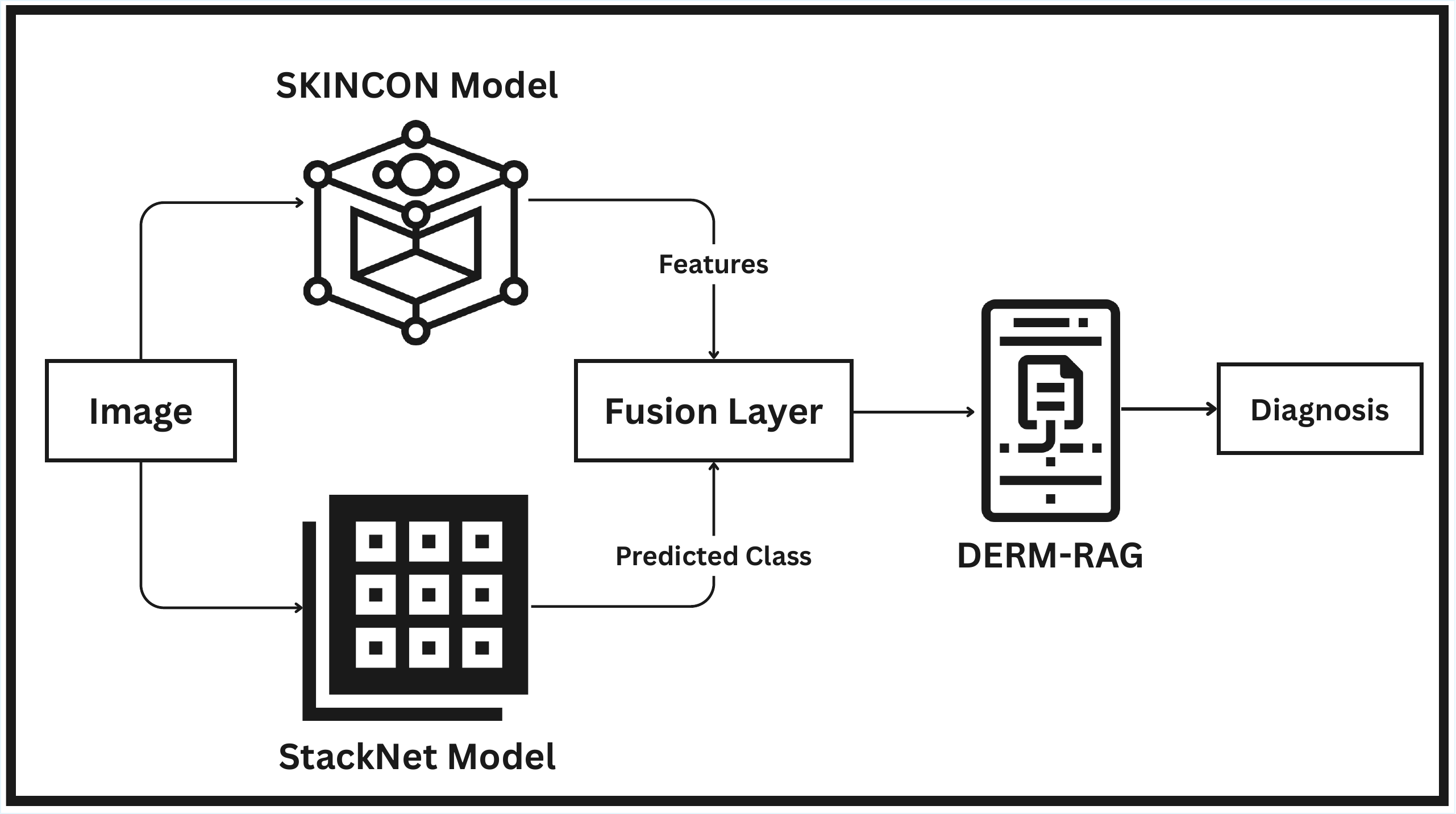}
  \captionof{figure}{DermETAS-SNA LLM Assistant System Pipeline}
\label{fig:system}
\end{figure}

As shown in Fig.\ref{fig:system}, in the first phase, we apply ETAS on the SKINCON dataset to obtain an optimal ViT architecture for multi-label classification, capable of capturing fine-grained dermatological features across diverse visual patterns. This architecture is then fine-tuned into a set of 23 class-specific binary classifiers using the DermNet dataset, enhancing diagnostic granularity and robustness, particularly for rare conditions often underrepresented in dermatological datasets.

In the second phase, the predictions from the visual models are forwarded to a RAG-based pipeline to generate personalized, clinically grounded explanations. Our knowledge source consists of curated content from standard dermatology textbooks, parsed using the LLaMA Parser \cite{lama_parser}, and embedded via the GTE Qwen2-1.5B Instruct model \cite{qwen}. QdrantDB \cite{qdrant} serves as the vector store for fast semantic retrieval, with Cohere’s Re-ranker-v3.5 \cite{cohere} refining document relevance prior to final synthesis.

We evaluated the response generations using four backends: GPT-4o \cite{gpt4o}, LLaMA 4 Maverick \cite{llama4}, Gemini 2.5 Pro \cite{gemini25pro}, and a multi-LLM ensemble fused through LLM-Blender's \cite{llmblender} PairRM and GenFuser modules to determine the most effective solution for our use case. This architecture enables trusted, context-sensitive responses with a reduced risk of hallucination. The pipeline is deployed in a functional prototype that supports real-time image input, symptom prompts, and secure diagnostic feedback.

\section{Evolutionary Transformer Architecture Search (ETAS)}

\begin{figure}[t]
\centering
  \includegraphics[width=0.72\textwidth]{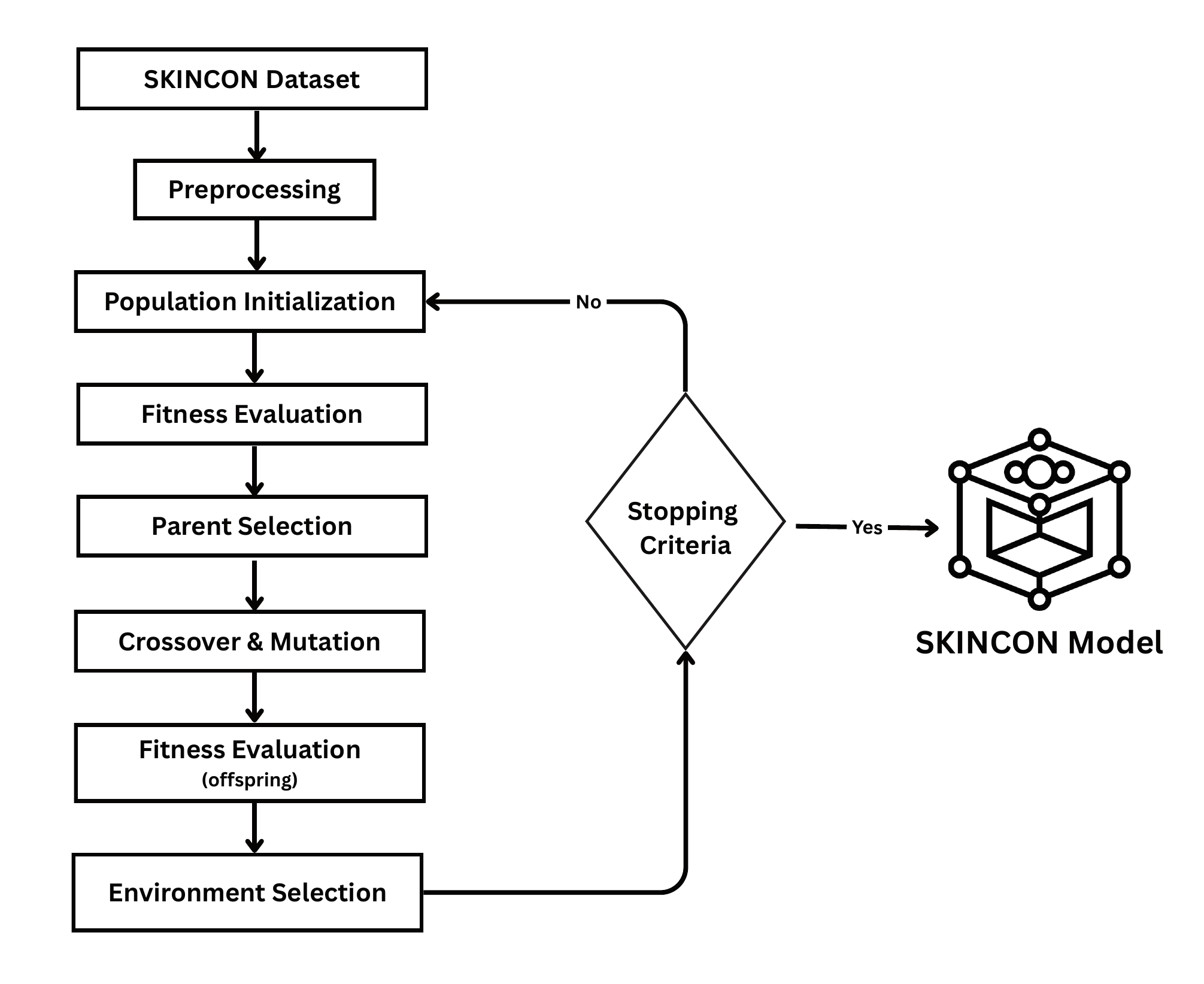}
  \captionof{figure}{ETAS Training Pipeline}
  \label{fig:etas}
\end{figure}

As shown in our ETAS training pipeline in Fig.\ref{fig:etas}, the genetic algorithm (GA) evolves ViT architectures through iterative selection. Our work builds upon and adapts the ENAS-FERNet framework proposed by Deng et al.~(2023) \cite{ENAS}, which employs Evolutionary Neural Architecture Search (ENAS) for facial expression recognition. We extend this approach to design an ETAS specifically tailored for optimizing ViT architectures for multi-label skin image classification using the SKINCON dataset. 

The goal of the framework is to find the ViT architecture that achieves the best fitness performance on a given image classification task. To apply GA in this context, we define a genetic representation of each ViT architecture candidate. The genetic operators work in conjunction with the evolutionary process shown in Fig.\ref{fig:etas}. The \textbf{crossover} operator combines architectural components from parent models, the \textbf{mutation} operator introduces random variations to explore new architectures, and the \textbf{selection} operator preserves the best-performing models for the next generation. This iterative process continues until the convergence criteria are met.
This evolutionary process helps in finding high performing ViT architectures that serve as our base or \emph{Step-1 model}, providing the foundation for further fine-tuning and application in subsequent modules of our framework.



The complete evolutionary process is formalized in Algorithm~\ref{alg:main_etas}, which describes the core optimization loop where ViT architectures are iteratively generated, evaluated, and selected across generations. Given that fitness evaluation (Algorithm 2) represents a computational bottleneck in this process, we implement a hash-based caching mechanism to store and reuse performance evaluations of previously encountered architectures, thereby significantly reducing redundant computation. To illustrate the mechanism of each phase in our ETAS framework, we use the specific example of optimizing architecture for the classification of melanoma throughout the following detailed explanation.

\begin{algorithm}[htbp]
\DontPrintSemicolon
\footnotesize
\SetAlgoLined
\caption{ETAS: Main Evolutionary Loop}
\label{alg:main_etas}
\KwIn{$max\_gens, pop\_size, p_{cross}, p_{mutate}$}
\KwOut{The ViT architecture with the highest fitness found.}
\BlankLine
\tcp{\textbf{Initialization: Randomly create the first generation}}
\BlankLine
$P \gets \text{empty list of Individuals}$\;
\For{$i \gets 1$ \KwTo $pop\_size$}{
    $I \gets \text{new Individual()}$\;
    $I.\text{architecture} \gets \text{GenerateRandomArchitecture()}$\ \tcp*{Random ViT layer config}
    $P.\text{append}(I)$\;
}
$P \gets \text{EvaluateFitness}(P)$ \tcp*{Train and assign fitness to initial population using Algorithm~\ref{alg:etas_fitness}}
\BlankLine
\tcp{\textbf{Evolutionary Loop: Run for max\_gens generations}}
\BlankLine

\For{$g \gets 1$ \KwTo $max\_gens$}{
    $O \gets \text{empty list of offspring}$\;
    \tcp{\textbf{Generate offspring via crossover and mutation}}

    \For{$i \gets 1$ \KwTo $pop\_size/2$}{
        $p_1, p_2 \gets \text{TournamentSelect}(P)$\ \tcp*{Select parents based on fitness}
        
        \eIf{$\text{random}() < p_{cross}$}{
            $c_1, c_2 \gets \text{SinglePointCrossover}(p_1, p_2)$ \tcp*{Cross over parents to produce new children architecture}
        }{
            $c_1, c_2 \gets p_1, p_2$ \tcp*{No crossover; use parent architectures directly}
        }
        \If{$\text{random}() < p_{mutate}$}{
            Mutate $c_1$ and $c_2$ by adding/removing a layer or changing a hyperparameter\;
        }
        \text{append $c_1$ and $c_2$ to $O$}
    }
    \BlankLine
    
    $O \gets \text{EvaluateFitness}(O)$\ \tcp*{Train children and assign fitness using Algorithm~\ref{alg:etas_fitness}}
    \BlankLine

    \tcp{\textbf{Environment Selection : Choose best candidates for next generation}}
    \BlankLine
    $P_{combined} \gets P \cup O$ \tcp*{Merge parents and offspring}
    $I_{best} \gets \text{Find best individual with highest fitness in } P_{combined}$\;
    $P_{next} \gets \text{RouletteWheelSelect}(P_{combined}, pop\_size)$ \tcp*{Selection based on fitness with high probability}
    \If{$I_{best} \notin P_{next}$}{
        Replace worst individual in $P_{next}$ with $I_{best}$ \tcp*{Elitism: preserve best solution}
    }
    $P \gets P_{next}$\;
}
\BlankLine
\Return{\text{Indidual with the highest fitness}(P)}\;
\end{algorithm}


\begin{algorithm}[htbp]
\DontPrintSemicolon
\SetAlgoLined
\SetKwProg{Fn}{Procedure}{}{}
\Fn{EvaluateFitness(Population)}{
    \For{$I$ \textbf{in} Population}{
        $key \gets \text{Hash}(I.\text{architecture})$ \tcp*{Unique key for caching}
        \eIf{$key$ is in cache}{
            $I.fitness \gets \text{cached\_value}[key]$ \tcp*{Use stored score if already evaluated}
        }{
            Train $I$ with 5-fold cross-validation \tcp*{Train the model on SKINCON dataset}
            $I.fitness \gets \text{average F1 score from training run}$ \tcp*{Evaluate performance}
            Store $I.fitness$ in cache with $key$ \tcp*{Cache for efficiency}
        }
    }
    \KwRet{Population}
}
\caption{Fitness Evaluation Procedure}
\label{alg:etas_fitness}
\end{algorithm}

\subsection{Mathematical Formulation}
\subsubsection{The "Individual" as a ViT Architecture:}
An "Individual" in the genetic algorithm is a complete Vision Transformer architecture. Its genetic code or "chromosome" is a set of hyperparameters that define the model: (1) \textbf{Chromosome}: A variable-length sequence of Transformer Layers and (2) \textbf{Genes}: The hyperparameters of each Transformer Layer, plus the total number of layers.

An individual \(I\) can be formally represented as a sequence of \(n\) transformer layer L configurations:
\begin{equation}
I = (L_1, L_2, \dots, L_n)
\label{eq:individual_representation}
\end{equation}
, where \(n\) is the number of transformer layers, a value between a defined minimum (e.g., 6) and maximum (e.g., 12).

Each layer \(L_i\) is a tuple of its specific hyperparameters (genes):
\begin{equation}
L_i = (h_i, m_i, d_i, p_i)
\label{eq:layer_definition}
\end{equation}

, where \(h_i\) denotes the number of attention heads in the \(i\)-th transformer layer, \(m_i\) is the dimension of the multilayer perceptron (MLP) in the \(i\)-th layer, \(d_i\) represents the dropout rate applied in the \(i\)-th layer, and \(p_i\) includes additional layer-specific parameters. The search space parameters and their values are given in Table \ref{tab:params}.

\begin{table}[!h]
\caption{GA Settings and ViT Hyperparameter Search Space}
\centering
\begin{tabular}{|l|l|l|}
\hline
\textbf{Category} & \textbf{Parameter} & \textbf{Value / Search Space} \\
\hline
\multicolumn{3}{|c|}{\textbf{Genetic Algorithm Settings}} \\
\hline
Population & Population Size & 5 \\
\hline
Operators & Crossover Probability & 0.8 \\
          & Mutation Probability & 0.2 \\
          & Mutation Type Probs & [0.7, 0.2, 0.1] for [Add, Remove, Modify] \\
\hline
\multicolumn{3}{|c|}{\textbf{Evolved ViT Hyperparameters (The Search Space)}} \\
\hline
Architecture & Transformer Layers & Integer in [6, 12] \\
\hline
Layer-specific & Attention Heads & \{8, 16\} \\
               & MLP Dimension & \{2048, 3072, 4096\} \\
               & Dropout Rate & Continuous in [0.1, 0.3] \\
\hline
\multicolumn{3}{|c|}{\textbf{Fixed ViT Hyperparameters}} \\
\hline
Dimensions & Hidden Dimension & 512 \\
           & Embedding Dimension & 768 \\
\hline
Input & Image Size & 224x224 \\
      & Patch Size & 16x16 \\
\hline
\end{tabular}
\label{tab:params}
\end{table}

\subsubsection{The Fitness Function:}
The objective of the GA is to maximize a fitness function. Here, the fitness of an individual is its \textbf{F1 score} after the corresponding ViT model has been trained and evaluated using 5-fold cross-validation.


\begin{equation}
\text{Fitness}(I) = \frac{1}{5} \sum_{k=1}^{5} \text{F1-Score}_k(\text{TrainAndEvaluate}(I))
\label{eq:fitness_function}
\end{equation}

The algorithm's goal is to find the individual \(I^*\) with the maximum fitness value of all possible architectures:
\begin{equation}
I^* = \arg\max_{I \in \mathcal{S}}(\text{Fitness}(I))
\label{eq:optimal_architecture}
\end{equation}


, where \textit{S} is a set of ViT architecture candidates. The choice of a relatively small population size ($pop\_size = 5$) and a limited number of generations ($max\_gens = 20$) represents a pragmatic trade-off between computational feasibility and exploratory sufficiency. The evolutionary search for optimal architectures is computationally expensive, as each individual candidate in the population must be trained and evaluated to assess its fitness. Given that the SKINCON dataset, while densely annotated, is of a moderate size (3,886 images), an extensive search with a large population was deemed unnecessary to find a strong local optimum. Our analysis indicated that a population of 5 individuals over 20 generations provided a sufficient diversity of architectures for the search to converge effectively on this dataset, without incurring prohibitive computational costs.


\subsection{Illustrative Example}
To demonstrate the application of our mathematical formulation, we present a concrete example of the ETAS pipeline optimizing an architecture for melanoma classification. The input image in Fig.\ref{fig:main} shows a sample dermoscopic image from the Melanoma class that would be processed through our system. The evolutionary process unfolds as follows:

\begin{itemize}
    \item[] \textbf{Step 1: Initialization (Algorithm~\ref{alg:main_etas}, lines 3-7)}
    \begin{itemize}
        \item[] Input: Fig.~\ref{fig:main} from Melanoma class
        \item[] Initial Architecture (Gen 1):
        \begin{equation}
        I_1 = [(8,2048,0.2), (8,2048,0.2), (16,3072,0.1)]
        \label{eq:init_arch}
        \end{equation}
        This architecture represents a 3-layer ViT with the gene configurations specified in Equation~\ref{eq:layer_definition}.
    \end{itemize}
    
    \item[] \textbf{Step 2: Fitness Evaluation (Algorithm~\ref{alg:etas_fitness})}
    \begin{itemize}
        \item[] The architecture undergoes training and evaluation, achieving:
        \begin{equation}
        \text{Fitness}(I_1) = 0.72
        \end{equation}
        This F1-score represents the performance metric defined in Equation~\ref{eq:fitness_function}.
    \end{itemize}
    \item[] \textbf{Step 3: Evolutionary Optimization (Algorithm~\ref{alg:main_etas}, lines 8-30)}
    \begin{itemize}

        \item[] Through 20 generations of selection, crossover, and mutation operations:
        \item[] Final Optimized Architecture (Generation 20):
        \begin{equation}
        I^* = [(16,4096,0.1), (16,3072,0.2), (16,3072,0.1)]
        \label{eq:final_arch}
        \end{equation}
        This architecture maximizes the objective function in Equation~\ref{eq:optimal_architecture}.
    \end{itemize}
    
    \item[] \textbf{Step 4: Feature Extraction}
    \begin{itemize}

        \item[] The optimized architecture extracts feature vector:
        \begin{equation}
        \mathbf{v} = [\text{Papule}, \text{Scale}, \text{Brown Hyperpigmentation}]
        \end{equation}
    \end{itemize}
\end{itemize}   

This feature vector \(\mathbf{v}\) serves as input to both StackNet for disease classification and DERM-RAG for evidence retrieval and explanation generation.

\section{StackNet Architecture}
\label{sec:stacknet}

\begin{figure}[t]
\centering
\includegraphics[width=0.70\textwidth]{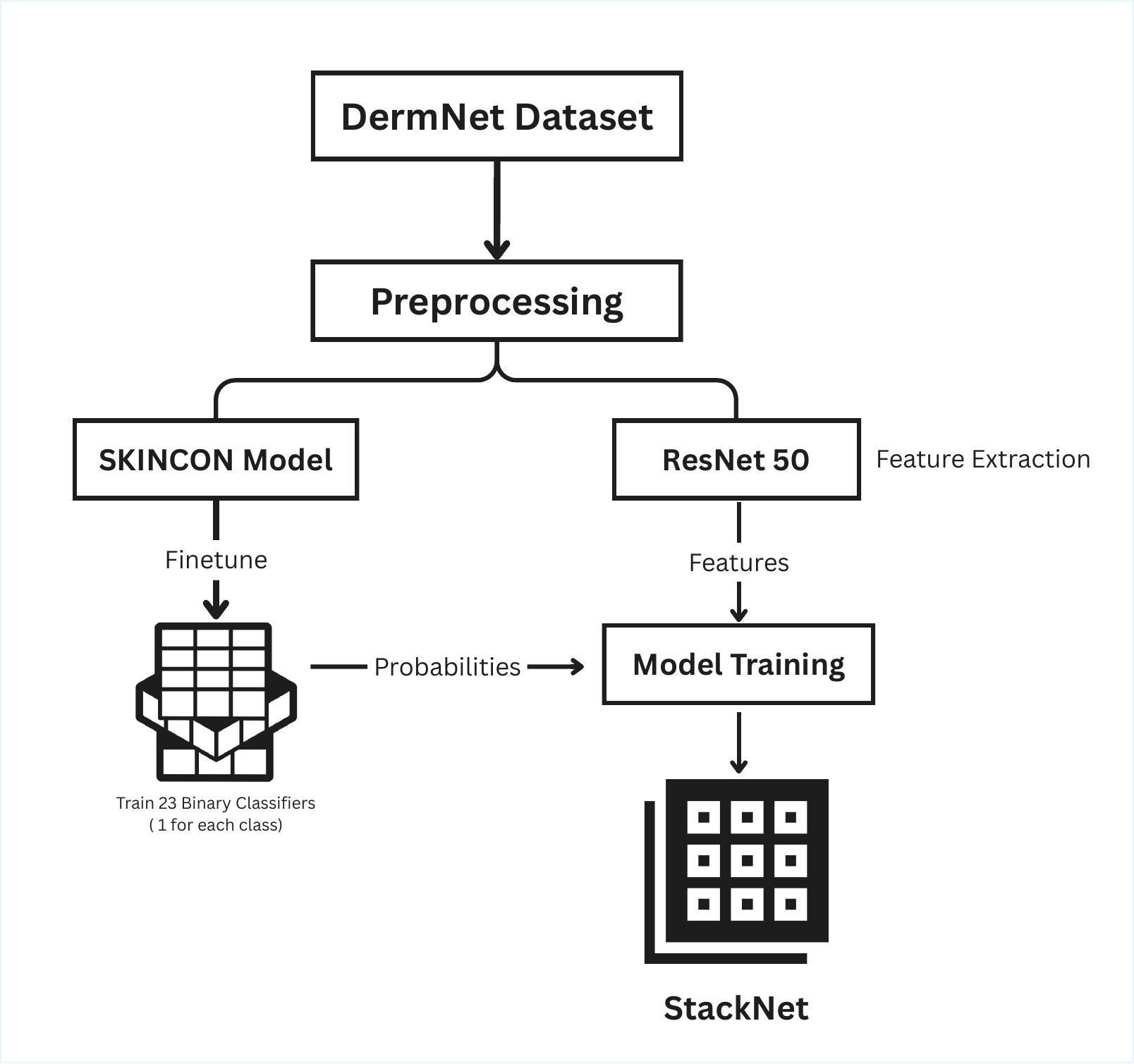}
\caption{StackNet Architecture.}
\label{fig:Stacknet}
\end{figure}

In the second module of our framework as shown in Fig.\ref{fig:Stacknet}, we build upon the best ViT architecture identified by ETAS on the SKINCON dataset. This model, originally optimized for multi-label classification of dermatological features, serves as the foundation for fine-tuning a suite of 23 binary classifiers, one dedicated to each disease class in the DermNet dataset.



To address key challenges in Dermatological AI, including poor performance on rare or underrepresented conditions and reduced diagnostic precision when using a single multi-class model, we employed a two-level model for skin cancer classification. The first level consists of training the 23 binary classifiers, allowing each model to focus on distinguishing a specific condition from all others. The second level involves a custom meta-classifier that synthesizes a complex set of features, including binary predictions, multi-scale deep features extracted from a pretrained model, and probability statistics to make a final, informed diagnosis.

\subsubsection{Stage 1: Binary Classifier Fine-Tuning:}

As outlined in Algorithm~\ref{alg:stacknet_stage1}, For each class \(c \in \{1, \dots, 23\}\), a dedicated binary classifier \(M_c\) is trained, by modifying its final classification layer of a ETAS-optimized ViT model to produce a single sigmoid-activated output, representing the probability of the image belonging to the target disease class. To optimize performance for each specific disease class, we employed two distinct fine-tuning strategies: (1) \textbf{Full Fine-tuning with Unfreezing (FU):} The entire pre-trained model was fine-tuned, allowing all weights to be updated for the new binary task; (2) \textbf{Gradual Layer Unfreezing (GU):} Fine-tuning commenced with only the final classification layers unfrozen. After each epoch, earlier layers were progressively unfrozen, allowing the model to adapt more stably.

A key aspect is the creation of a \textbf{balanced one-vs-all dataset} \(D_c\) for each class. The training objective is to find the optimal hyperparameters \(h^*\) (learning rate, momentum, batch size) from a predefined grid \(H\) by maximizing the accuracy using 5-fold and 10-fold cross-validation techniques. The final model \(M_c^*\) produces a probability \(p_c(x)\) for an input image \(x\).

\subsubsection{Stage 2: Meta-Classifier Feature Engineering:}
As detailed in Algorithm~\ref{alg:stacknet_stage2}, the second level employs a meta-classifier \(M_{\text{meta}}\) that predicts the skin disease class based on a comprehensive feature vector formed by concatenating the following three components:

\begin{itemize}
    \item \textbf{Probability Vector \(\mathbf{P}(x)\)}: Raw probabilities from the 23 binary classifiers.
    \begin{equation}
    \mathbf{P}(x) = [p_1^*(x), p_2^*(x), \dots, p_{23}^*(x)] \in \mathbb{R}^{23}
    \label{eq:prob_vector}
    \end{equation}

    \item \textbf{Multi-Scale Deep Feature Vector \(\mathbf{D}_{\text{multi}}(x)\)}: Features extracted from four layers of a pretrained ResNet-50 \cite{resnet-50} and concatenated to form a hierarchical representation:
    \begin{equation}
    \mathbf{D}_{\text{multi}}(x) = \bigoplus_{i=1}^{4} \text{GlobalAvgPool}(\text{ResNet}_{\text{L}_i}(x)) \in \mathbb{R}^{N}
    \label{eq:deep_feature_vectors}
    \end{equation}

    \item \textbf{Probability Statistics Vector \(\mathbf{S}(x)\)}: Summary statistics derived from \(\mathbf{P}(x)\) that capture distributional characteristics:

    \begin{equation}
    \mathbf{S}(x) = [\mu(\mathbf{P}), \sigma(\mathbf{P}), \mu(\text{top3}), (\text{top1} - \text{top3})] \in \mathbb{R}^{4}
    \label{eq:statistics_vector}
    \end{equation}

\end{itemize}

The final input to the meta-classifier is the concatenated feature vector:

\begin{equation}
\mathbf{F}(x) = \mathbf{P}(x) \oplus \mathbf{D}_{\text{multi}}(x) \oplus \mathbf{S}(x)
\label{eq:feature_vector}
\end{equation}
The meta-classifier is implemented as a custom 1D CNN and trained using \textbf{Focal Loss} to effectively address class imbalance and focus on hard examples.




\begin{algorithm}[htbp]
\DontPrintSemicolon
\SetAlgoLined
\caption{Stage 1: Binary Model Training (Per Class)}
\label{alg:stacknet_stage1}
\KwIn{Base model \(\text{ViT}_{\text{ETAS}}\), Dermnet dataset \(D\), Class \(c\), Hyperparameter Grid \(H\), Fold \(F\)}
\KwOut{The single best-performing binary model \(M_c^*\) for class \(c\)}
\BlankLine
$D_c^{\text{bal}} \gets \text{CreateBalancedBinaryDataset}(D, c)$ \tcp{Balance dataset for class $c$ vs all}
$best\_global\_f1 \gets 0$\;
$best\_global\_model \gets \text{None}$ \tcp{Track best model}
\For{k \textbf{in} Fold\(F\)}{
    \For{params \textbf{in} Hyperparameter Grid \(H\)}{ 
        fold\_f1s $\gets []$\;
        \For{fold \textbf{in} k-Fold CV split of $D_c^{\text{bal}}$}{
            model $\gets \text{Load}(\text{ViT}_{\text{ETAS}})$ \tcp{Load base ViT-ETAS}
            model.classifier $\gets \text{nn.Linear}(512, 1)$ \tcp{Replace last layer for binary classification}
            \text{FineTune}(model, params, fold.train\_data) \tcp{Train model}
            f1 $\gets \text{Evaluate}(model, fold.val\_data)$\;
            fold\_f1s.append(f1)\;
        }
        \If{Average(fold\_f1s) $>$ best\_global\_f1}{ \tcp{Check if current model is best so far}
            best\_global\_f1 $\gets$ Average(fold\_f1s)\;
            best\_global\_model $\gets$ \text{best model saved during this run}\;
        }
    }
}
\Return{best\_global\_model} \tcp{Return best model for class $c$}
\end{algorithm}
\begin{algorithm}[htbp]
\DontPrintSemicolon
\SetAlgoLined
\caption{Stage 2: Full Prediction Pipeline with Feature Engineering}
\label{alg:stacknet_stage2}
\KwIn{Input Image \(x\), Binary Models\(\{M_c^*\}\), Pretrained ResNet50, Meta-Classifier \(M_{\text{meta}}\)}
\KwOut{Final Class Prediction}
$P_{vec} \gets [M_1^*(x), \dots, M_{23}^*(x)]$ \tcp{23-dim probability vector}
$D_{multi} \gets \text{ExtractMultiLayerResNetFeatures}(x)$ \tcp{Extract deep features from ResNet layers}
$S_{vec} \gets \text{CalculateProbabilityStats}(P_{vec})$ \tcp{Compute statistical features (mean, std, etc.)}
$F_{final} \gets \text{Concatenate}(P_{vec}, D_{multi}, S_{vec})$ \tcp{Combine all features into final vector}
prediction $\gets M_{\text{meta}}(F_{final})$ \tcp{Predict final class using meta-classifier}
\Return{prediction}\;
\end{algorithm}


The input image from Fig.\ref{fig:main} and optimized ViT architecture are processed through StackNet:

\begin{itemize}
    \item[] \textbf{Step 1: Binary Classifier Training (Algorithm~\ref{alg:stacknet_stage1})}
    \begin{itemize}
        \item[] Balanced dataset: 300 Melanoma vs 300 non-Melanoma images
        \item[] Fine-tuned using ViT($I^*$) with sigmoid head
        \item[] Best model $M_{mel}^*$ predicts:
        \begin{equation}
        p_{\text{mel}}(x) = \textbf{0.91}
        \label{eq:melanoma_prob}
        \end{equation}
    \end{itemize}

    \item[] \textbf{Step 2: Probability Vector Construction (Algorithm~\ref{alg:stacknet_stage2}, line 2)}
    \begin{equation}
    \mathbf{P}(x) = [p_1(x), p_2(x), \dots, p_{23}(x)] \in \mathbb{R}^{23}
    \label{eq:prob_vector}
    \end{equation}

    \item[] \textbf{Step 3: Deep Feature Extraction (Algorithm~\ref{alg:stacknet_stage2}, line 3)}
    \begin{equation}
    \mathbf{D}_{\text{multi}}(x) = \bigoplus_{i=1}^4 \text{GAP}(\text{ResNet}_{\text{L}_i}(x)) \in \mathbb{R}^{2048}
    \label{eq:deep_features}
    \end{equation}

    \item[] \textbf{Step 4: Statistical Features (Algorithm~\ref{alg:stacknet_stage2}, line 4)}
    \begin{equation}
    \mathbf{S}(x) = \begin{bmatrix}
    \mu(\mathbf{P}) \\
    \sigma(\mathbf{P}) \\
    \mu(\text{top3}(\mathbf{P})) \\
    \text{max}(\mathbf{P}) - \mu(\text{top3}(\mathbf{P}))
    \end{bmatrix} = \begin{bmatrix}
    0.18 \\
    0.22 \\
    0.72 \\
    0.13
    \end{bmatrix}
    \label{eq:stats}
    \end{equation}

    \item[] \textbf{Step 5: Meta-Classification (Algorithm~\ref{alg:stacknet_stage2}, lines 5-6)}
    \begin{equation}
    \mathbf{F}(x) = \mathbf{P}(x) \oplus \mathbf{D}_{\text{multi}}(x) \oplus \mathbf{S}(x) \in \mathbb{R}^{2075}
    \label{eq:final_feature}
    \end{equation}
    \begin{equation}
    \text{Final Prediction} = \text{argmax}(\text{MLP}(\mathbf{F}(x))) = \boxed{\text{Melanoma (94.5\%)}}
    \label{eq:final_pred}
    \end{equation}
\end{itemize}

\vspace{-2mm}
\section{DERM-RAG - Diagnostic Explanation and Retrieval Model for Dermatology Pipeline}

\begin{figure}[t]
\centering
\includegraphics[width=\textwidth]{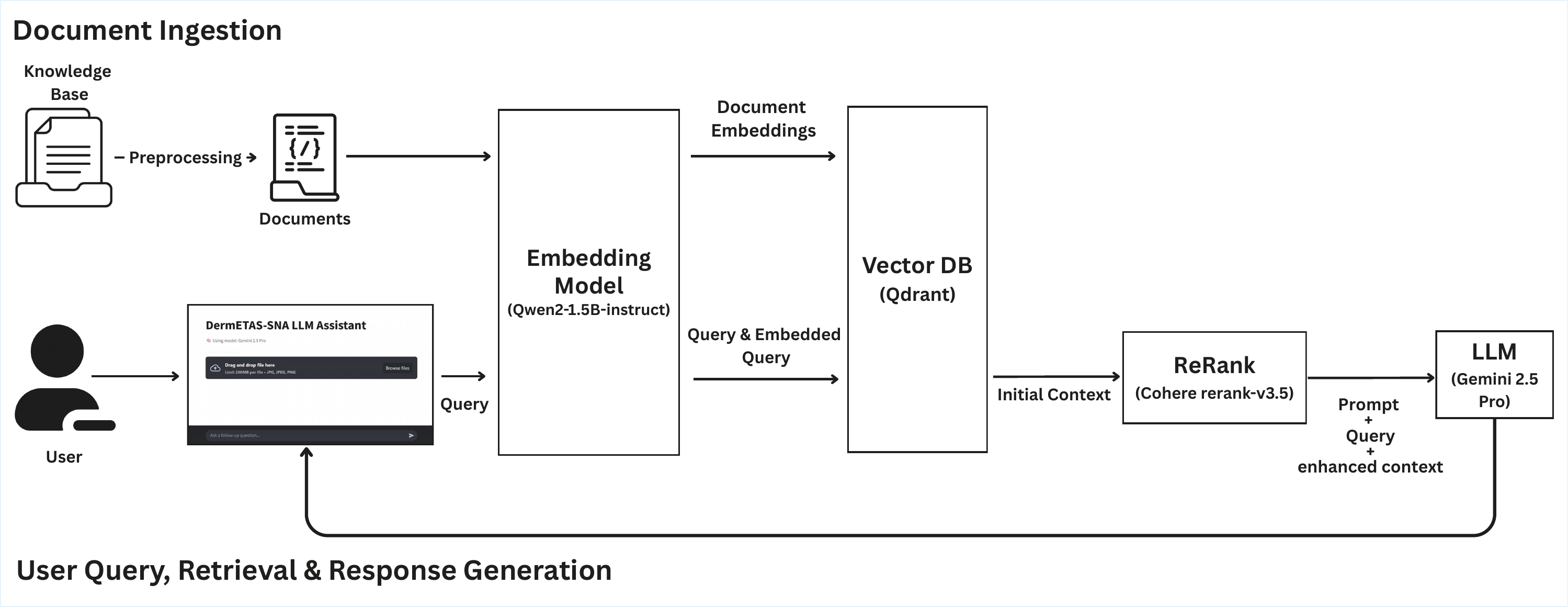}
  \captionof{figure}{DERM-RAG Pipeline}
  \label{fig:dermrag}
\end{figure}

In the DERM-RAG Module, we focus on giving users clear and trustworthy answers using \textbf{Retrieval-Augmented Generation (RAG)}. As shown in Fig.\ref{fig:dermrag}, first we gathered around 5,000 pages of trusted skin disease information from well-known materials like \textit{Habif’s Clinical Dermatology}, \textit{Rook’s Textbook of Dermatology}, \textit{Fitzpatrick Dermatology}, and a few others. This information is cleaned and broken into small, meaningful chunks using \textbf{LLaMA Parser}. For example, one chunk might explain \textit{``what melanoma is''} and another might talk about \textit{``treatment options''}.

We then use \textbf{Qwen 1.5B} Embedding Model to convert each chunk into embeddings. These embeddings are stored in an efficient semantic search engine called \textbf{QdrantDB}, allowing quick retrieval of the most relevant information when a user submits a question. For example, if a user uploads an image, and the system predicts \textbf{Melanoma} and also detects features such as \textit{Papule} and \textit{Brown (Hyperpigmentation)}. Based on this, the system constructs a user-friendly question like:


\begin{quote}
\textit{``The image analysis detected the following skin features: Papule, Brown (HyperPigmentation). The most likely diagnosis based on our model is: Melanoma. Please explain what this diagnosis usually means, common symptoms, and first-line treatments in simple terms?''}
\end{quote}




This question is then embedded using \textbf{Qwen 1.5B}, and the most relevant chunks are retrieved from the knowledge base. Example results might include: (1) \textit{``Melanoma is a serious form of skin cancer that starts in pigment cells.''}; (2) \textit{``Treatments include surgery, immunotherapy, and targeted drugs like BRAF inhibitors.''}.


These chunks are passed through a \textbf{Cohere Re-Ranker}, which reorders them based on relevance. Subsequently, one or more advanced LLMs, such as \textbf{Gemini 2.5 Pro}, \textbf{GPT-4o}, or \textbf{LLaMA 4 Maverick} generate a friendly and informative response. For example:

\begin{quote}
\textit{``Melanoma is a dangerous skin cancer that comes from pigment-producing cells. It can be treated with surgery if caught early. In later stages, treatments like immunotherapy or targeted drugs may be used. Early diagnosis is very important for a better outcome.''}
\end{quote}

To ensure these answers are accurate and safe, we involve \textbf{dermatologists} in an expert evaluation step. Multiple responses to the same question are presented to them, and each is graded on a scale of \textit{Strongly Agree, Agree, Neutral, Disagree, Strongly Disagree}. For example, if three dermatologists rate a response as \textit{Strongly Agree}, the answer is considered highly reliable. Through this process, our system integrates curated medical knowledge, semantic search, and advanced AI to deliver accurate and comprehensible answers validated by domain experts.

\section{Experimental Evaluations and Domain-Expert Assessment}
\subsection{Experimental Evaluations}




The evolutionary search in the ETAS framework ran for 20 generations with a population size of 5. Each candidate architecture was evaluated using 5-fold cross-validation, with the average performance determining selection for the next generation.
After evolution, the best-performing model was a 12-layer transformer, trained for 50 epochs, achieving the highest average F1-score of \textbf{0.716} across all folds. To address the severe class imbalance in the DermNet dataset, we trained 23 specialized binary classifiers, one for each disease class, using the fine-tuning strategies detailed in Section~\ref{sec:stacknet}. Extensive experimentation was conducted with hyperparameters such as learning rate, momentum, and batch size using both 5-fold and 10-fold cross validation to determine the best-performing model for each class and selected for comparison against SkinGPT-4 as a baseline. Each binary models was assessed using the performance metrics such as Accuracy, Precision, Recall, F1-Score, and Matthews Correlation Coefficient (MCC).



The best models were selected based on both training strategies and 5-fold and 10-fold cross-validation results. Complete class-wise performance metrics are presented in Table \ref{tab:binary_models_performance}.
Among the seven classes suggested by domain experts, several showed particularly strong performance:
\begin{enumerate}
    \item Although Melanoma is not among the seven classes, \textbf{Melanoma} achieved the highest F1 score (74.16\%) by full unfreezing with 10-fold CV, demonstrating the effectiveness of our approach for critical diagnoses.
    \item \textbf{Atopic Dermatitis} and \textbf{Hair Loss} both exceeded 70\% F1-score, indicating robust performance for common conditions.
    \item \textbf{Actinic Keratosis} and \textbf{Seborrheic Keratoses} showed moderate performance (62.38\% and 65.08\% F1-score respectively), likely due to visual similarity between these conditions.
    \item \textbf{Psoriasis} and \textbf{Warts} had lower performance (44.52\% and 55.74\%), suggesting these classes may benefit from additional training data or architectural adjustments.
\end{enumerate}

\begin{table}[!h]
\caption{Binary Classifier Performance}
\centering
\begin{tabular}{@{}l l c c c@{}}
\toprule
\textbf{Class} & \textbf{Best Model} & \textbf{Accuracy} & \textbf{F1 Score} & \textbf{MCC} \\
\midrule
Acne and Rosacea       & GU - 10 Fold   & 0.5819 & 0.5956 & 0.1642 \\
Actinic Keratosis      & GU - 5 Fold    & 0.5938 & 0.6238 & 0.1899 \\
Atopic Dermatitis      & FU - 5 Fold    & 0.6707 & 0.7362 & 0.3932 \\
Bullous Disease        & GU - 10 Fold   & 0.6018 & 0.5909 & 0.2038 \\
Cellulitis Impetigo    & GU - 10 Fold   & 0.6027 & 0.6375 & 0.2094 \\
Contact Dermatitis     & GU - 5 Fold    & 0.6000 & 0.6389 & 0.2048 \\
Eczema                 & GU - 10 Fold   & 0.6133 & 0.6490 & 0.2314 \\
Exanthems              & GU - 10 Fold   & 0.6089 & 0.6457 & 0.2227 \\
Fungal Infections      & GU - 10 Fold   & 0.5446 & 0.3537 & 0.1106 \\
Hair Loss              & FU - 5 Fold    & 0.6083 & 0.7117 & 0.3107 \\
Herpes                 & FU - 5 Fold    & 0.6029 & 0.6824 & 0.2377 \\
Infestations \& Bites  & GU - 5 Fold    & 0.5324 & 0.4599 & 0.0673 \\
Light Diseases         & GU - 5 Fold    & 0.5804 & 0.6000 & 0.1616 \\
Lupus                  & FU - 5 Fold    & 0.7048 & 0.6961 & 0.4102 \\
Melanoma               & FU - 10 Fold   & 0.7026 & 0.7416 & 0.4250 \\
Nail Fungus            & GU - 5 Fold    & 0.6820 & 0.6937 & 0.3651 \\
Psoriasis              & GU - 10 Fold   & 0.5114 & 0.4452 & 0.0234 \\
Seborrheic Keratoses   & GU - 10 Fold   & 0.6356 & 0.6508 & 0.2722 \\
Systemic Diseases      & FU - 5 Fold    & 0.6053 & 0.6591 & 0.2219 \\
Urticaria Hives        & FU - 10 Fold   & 0.6792 & 0.7258 & 0.3811 \\
Vascular Tumors        & FU - 5 Fold    & 0.5865 & 0.6578 & 0.1904 \\
Vasculitis             & FU - 10 Fold   & 0.5667 & 0.6894 & 0.2177 \\
Warts                  & GU - 10 Fold   & 0.6029 & 0.5574 & 0.2104 \\
\bottomrule
\end{tabular}
\label{tab:binary_models_performance}
\end{table}
 The results demonstrate that specialized binary classifiers provide more reliable diagnoses than a unified model approach.
However, due to class overlaps and image duplication, direct aggregation of predictions from the binary classifiers caused misclassifications. For the meta classifier, we evaluated several CNN configurations and found the best results with a model containing three fully connected layers (1024, 512, 256 units), trained with a learning rate of 0.0005 and batch size of 16. To benchmark our system, we implemented SkinGPT-4 by adapting its base model trained first on SKINCON (20 epochs), then fine-tuned on DermNet (additional 20 epochs). Table~\ref{tab:meta_classifier_performance} compares the performance of StackNet against SkinGPT-4 across key evaluation metrics. The comparative analysis reveals several advantages of the StackNet architecture:


\begin{enumerate}
    \item \textbf{Superior Recall and F1-score:} StackNet's most prominent improvement is in \textbf{Recall} (55.29\% vs. 46.83\%), representing an increase by approximately 18.06\%. This indicates that our model is better at identifying true positives and reducing false negatives, critical in medical diagnostics where missing a disease (e.g., melanoma) can have severe consequences. The harmonized F1-score increased by approximately 16.06\%, improving from 48.51\% to 56.30\%, indicating a more robust overall performance.


    \item \textbf{Enhanced Classification Quality:} The higher Matthews Correlation Coefficient (MCC) of 0.57 compared to 0.50 signifies a better overall quality of binary classifications, especially important given the imbalanced nature of the DermNet dataset.

    \item \textbf{Architectural Advantages:} The performance gap can be attributed to the dual-level design of StackNet. SkinGPT-4 uses a single model for all 23 classes, which can struggle with inter-class similarities and data imbalance. In contrast, StackNet’s first-level binary classifiers specialize in individual conditions, learning highly discriminative features. A second-level meta-classifier then synthesizes these expert outputs with deep features and statistical context, yielding more nuanced and accurate predictions. This design is particularly effective for rare or visually similar conditions that are often misclassified by monolithic models.
\end{enumerate}

\begin{table}[t]
\caption{Performance Comparison between StackNet and SkinGPT-4}
\centering
\begin{tabular}{@{}l|c|c@{}}
\toprule
\textbf{Metric} & \textbf{StackNet} & \textbf{SkinGPT-4} \\
\midrule
Accuracy(\%)  & 59.89 & 52.92 \\
Precision(\%) & 59.25 & 54.57 \\
Recall(\%)    & 55.29 & 46.83 \\
F1 Score(\%)  & 56.30 & 48.51 \\
MCC       & 0.57 & 0.50 \\
\bottomrule
\end{tabular}
\label{tab:meta_classifier_performance}
\end{table}

\vspace{-5mm}





\subsection{Domain Expert Assessment}

\noindent We conducted a clinical evaluation of all models integrated into the DermETAS-SNA LLM pipeline. Following discussions with experts in the field, seven prominent skin conditions were chosen for comprehensive clinical assessment: Warts, Actinic Keratosis, Eczema, Seborrheic Keratoses, Psoriasis, Nail Fungus, and Hair Loss. The following user-centered evaluation questions were used to assess the generated responses: (1) Can you explain my diagnosis in detail, and how certain is it that I have skin disease?; (2) What are my treatment options for my diagnosis and possible side effects?; (3) What factors could influence my prognosis?; (4) How will the treatment affect my daily activities and overall quality of life?; (5) What kind of follow-up care and monitoring will be required after treatment? 

The responses were evaluated using the following six evaluation criteria: (1) The diagnosis is correct or relevant; (2) The explanation is informative; (3) Suggestions are useful; (4) Can help doctors with diagnosis; (5) Helps patients understand better; (6) Willingness to use this model.

\begin{figure}[h!]
    \centering
    \includegraphics[width=0.70\textwidth]{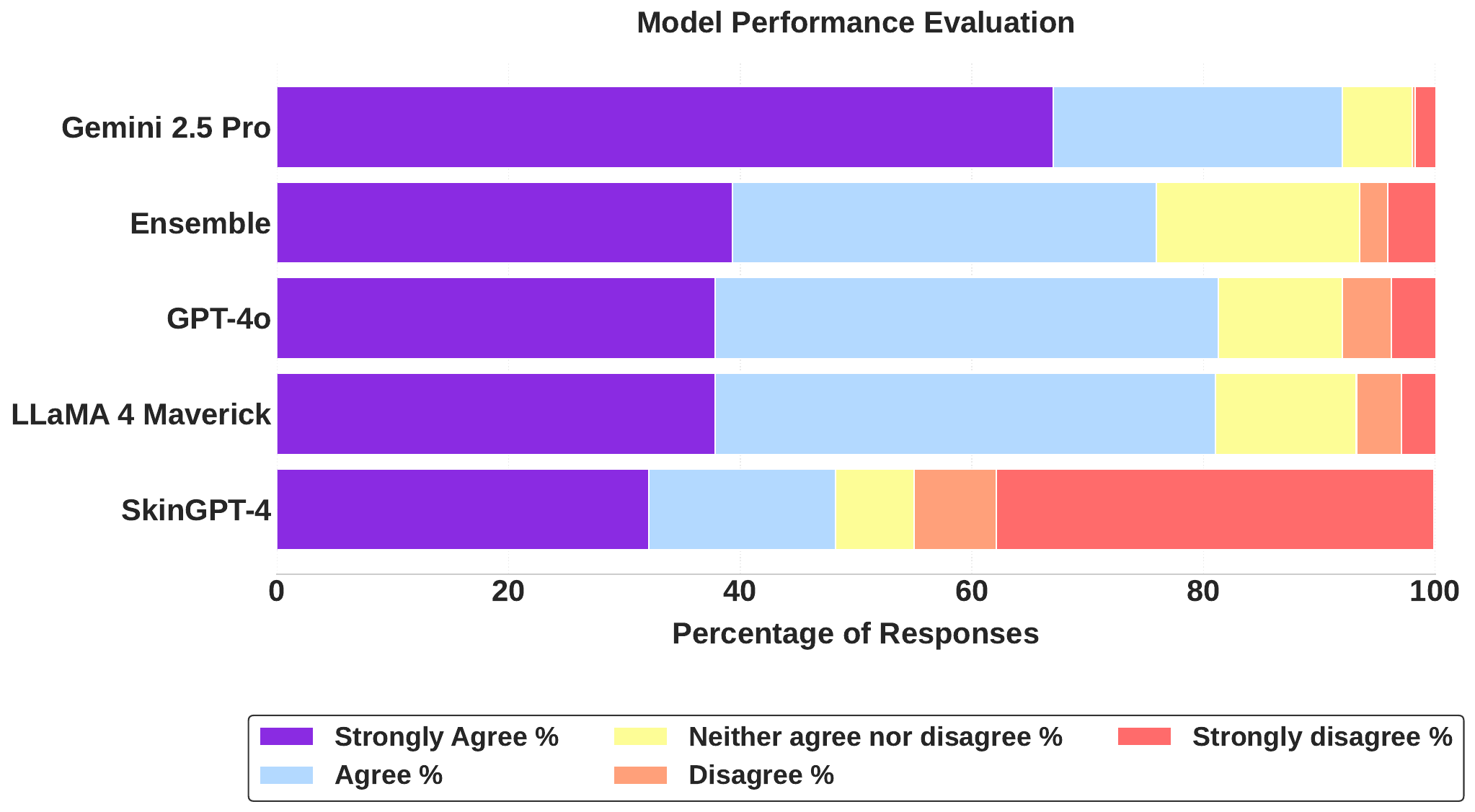} 
    \caption{Comparative Performance Evaluation of Models Showing Dermatologist's Agreement Levels with Model-Generated Responses.}
    \label{fig:performance}
\end{figure}

The responses are rated on a 5-point Likert scale for each criteria: Strongly Agree, Agree, Neutral, Disagree, Strongly Disagree. The aggregated expert ratings are shown in Table~\ref{tab:final_performance} and visualized in Fig.\ref{fig:performance}. Below are the critical findings from the evaluation results:




\begin{table}[t]
\caption{Clinical Agreement Rates between Dermatologists and AI Models.}
\centering
\renewcommand{\arraystretch}{1.0}

\begin{tabular}{
    @{}>{\centering\arraybackslash}m{3.0cm} 
      >{\centering\arraybackslash}m{1.8cm} 
      >{\centering\arraybackslash}m{1.4cm} 
      >{\centering\arraybackslash}m{2.0cm} 
      >{\centering\arraybackslash}m{1.6cm} 
      >{\centering\arraybackslash}m{1.7cm}@{}
}
\toprule
\textbf{Model} & 
\textbf{Strongly Agree (\%)} & 
\textbf{Agree (\%)} & 
\textbf{Neither Agree nor Disagree (\%)} & 
\textbf{Disagree (\%)} & 
\textbf{Strongly Disagree (\%)} \\
\midrule
Gemini 2.5 Pro         & 67.0 & 25.0 & 6.0  & 0.3 & 1.8 \\
LLM Blender - Ensemble & 39.3 & 36.6 & 17.6 & 2.4 & 4.2 \\
GPT-4o                 & 37.8 & 43.5 & 10.7 & 4.2 & 3.9 \\
LLaMA 4 Maverick       & 37.8 & 43.2 & 12.2 & 3.9 & 3.0 \\
SkinGPT-4              & 32.1 & 16.1 & 6.8  & 7.1 & 37.8 \\
\bottomrule
\end{tabular}
\label{tab:final_performance}
\end{table}



\begin{enumerate}
    \item \textbf{Superior Performance of Gemini 2.5 Pro with RAG:} The Gemini 2.5 Pro model, enhanced with our curated medical text and re-ranking pipeline (DERM-RAG), achieved the highest expert consensus with a 92\% agreement rate (67.0\% Strongly Agree + 25.0\% Agree). This underscores the value of grounding LLM responses in verified, authoritative sources. By retrieving and synthesizing content from dermatology textbooks, DERM-RAG mitigates hallucination and delivers accurate, contextually relevant explanations aligned with vision model outputs.


    \item \textbf{The Hallucination Problem in SkinGPT-4:} In stark contrast, SkinGPT-4 showed substantially lower agreement (48.2\%) and the highest rate of strong disagreement (37.8\%). This performance gap highlights a fundamental limitation: without integration to a verified knowledge base, even powerful multimodal LLMs are prone to generating plausible but incorrect or generalized information.

    \item \textbf{Performance of Other Foundational Models:} The other foundational models (GPT-4o, LLaMA 4 Maverick) and the Ensemble approach performed comparably, with overall agreement rates clustering around 80\%. This suggests that while they possess strong inherent capabilities, they still lack the domain-specific grounding necessary for high-stakes medical dialogue. The ensemble method did not outperform the best individual model (Gemini 2.5 Pro), indicating that simply blending outputs is less effective than a deeply integrated RAG approach for this specialized domain.
    
\end{enumerate}

\section{Proof of Concept Prototype}

The prototype user interface (UI) was developed using \texttt{Streamlit} and deployed on WPI’s Academic \& Research Computing cluster using an Apple Mac Studio (M2 Max, 12-core CPU, 96 GB unified memory, 38-core GPU). The stack included Python 3.10, Streamlit 1.25.0, and GPU-accelerated inference via Apple’s Metal API. 

\begin{figure}[!h]
  \centering
  \includegraphics[width=0.80\textwidth]{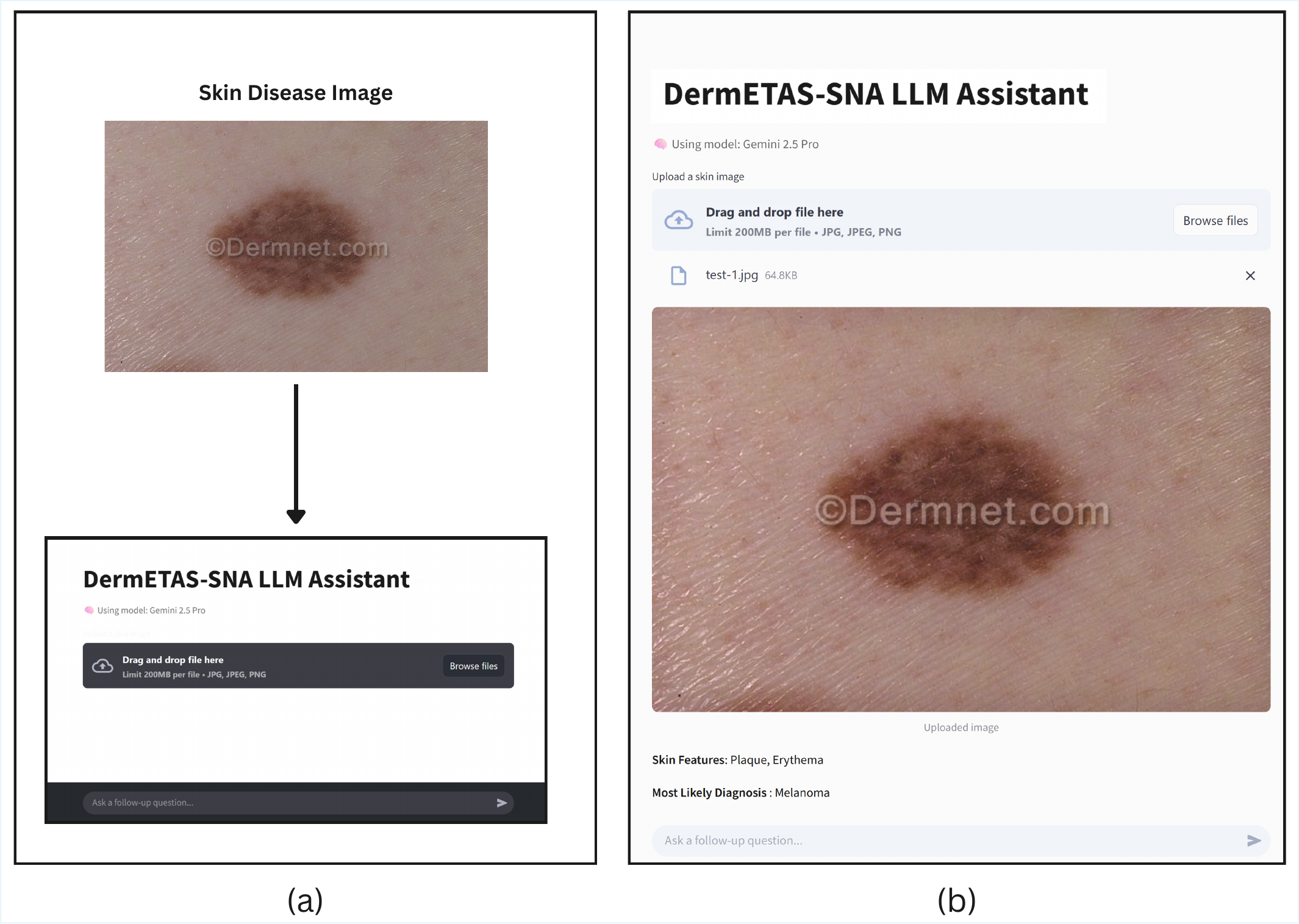}
  \vspace{0.5em} 
  \includegraphics[width=0.80\textwidth]{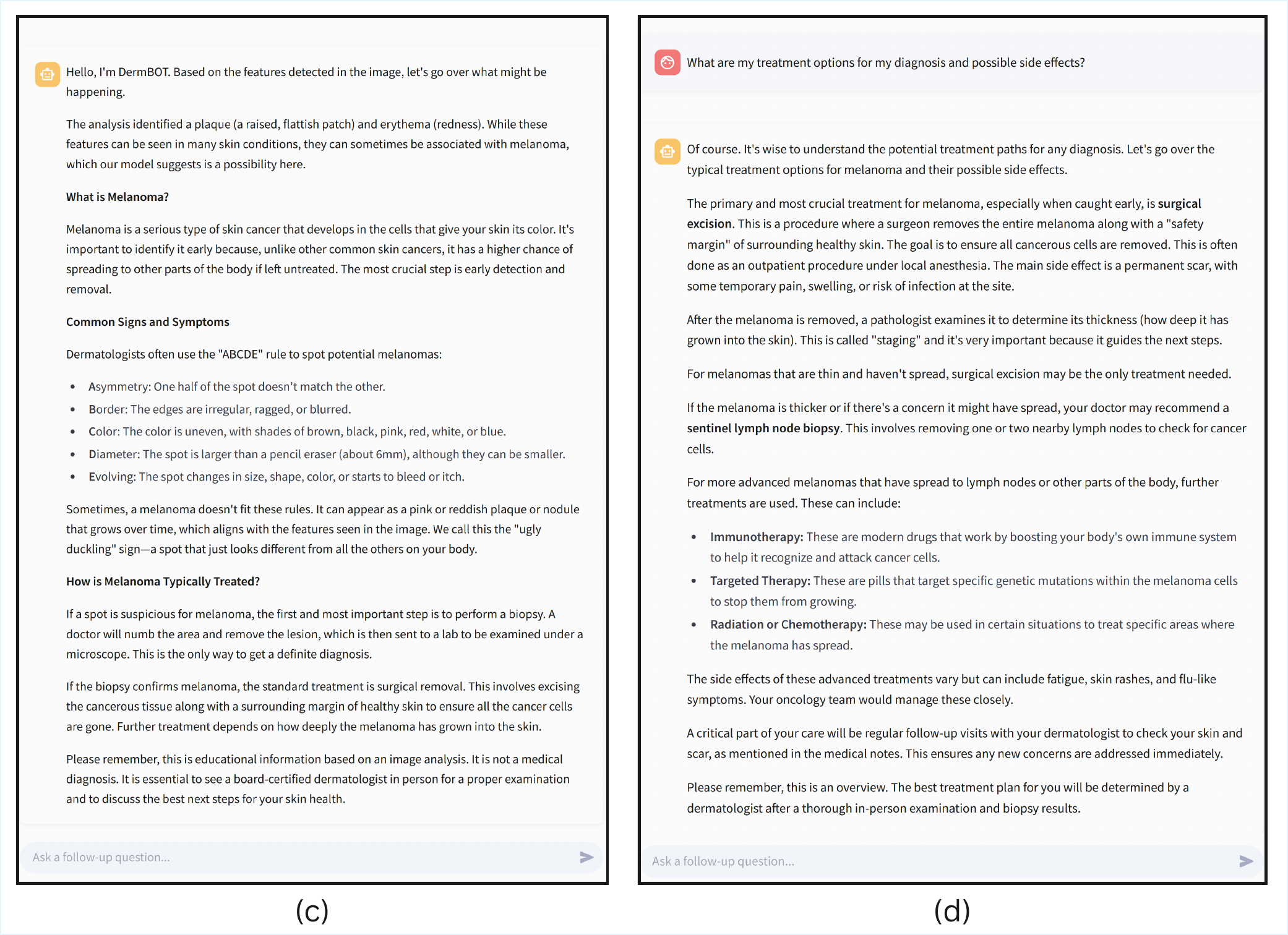}
  \caption{DermETAS-SNA Assistant Prototype Interface
  }
  \label{fig:dermbotprototype}
\end{figure}

Fig.\ref{fig:dermbotprototype} demonstrates the prototype's workflow: (a) users can upload dermatological images through an intuitive interface, (b) the system processes images through our ETAS and StackNet models to extract visual features and generate predictions, (c) the DERM-RAG module retrieves relevant information from medical knowledge bases to provide comprehensive diagnostic explanations, and (d) users engage in interactive dialogues about treatment options and side effects, highlighting multimodal conversational capabilities.

\section{Conclusions and Future Work}

This study demonstrates the potential of combining LLMs with classification and ensemble learning techniques, supported by domain expert evaluation, for effective skin disease diagnosis. Skin diseases have become one of the most prevalent health challenges globally, and our proposed AI assistant, \textit{DermETAS-SNA}, is designed to assist dermatologists by offering preliminary diagnostic support. Our key contributions include: (1) Development of an Evolutionary Transformer Architecture Search (ETAS) framework tailored for dermoscopic image analysis. (2) Implementation of a one-vs-all binary classification strategy to address the class imbalance present in the dataset. (3) Integration of domain-specific medical knowledge into LLM using retrieval-augmented generation (RAG), allowing tailored and context-aware responses.(4) Extensive experimental evaluations on 23 disease categories demonstrating a considerable improvement (16.06\% increase in F1-score) over the SkinGPT-4 baseline. (5) Evaluation of generated outputs through feedback from board-certified dermatologists to ensure clinical reliability and relevance. (6) Proof-of-concept prototype that fully integrates our DermETAS-SNA LLM into our AI assistant to demonstrate its practical feasibility for real-world clinical and educational applications.

We acknowledge that the absolute performance (56.3\% F1-score on a complex 23-class task) indicates a non-negligible error rate, positioning our system as an assistive tool for triage and differential diagnosis generation, not a replacement for clinical expertise. Clinically, the system reduces diagnostic wait times and enhances doctor-patient communication through explainable AI provided by our DERM-RAG module. Nonetheless, limitations remain. (1) Models were trained and evaluated on a standard stratified split of the DermNet dataset. While techniques like stratified splits and regularization were employed to mitigate overfitting in the one-vs-all classifiers, future work will involve rigorous cross-dataset validation on independent, publicly available benchmarks such as HAM10000 and ISIC to assess performance across diverse skin tones and imaging conditions. (2) As with many dermatological AI studies, demographic and phenotypic diversity is incompletely captured; a key goal is to train on larger, multi-source, ethically sourced datasets to ensure equitable performance.(3) Expand the RAG pipeline's knowledge base by incorporating authoritative medical sources (e.g., MedlinePlus, UMLS, UpToDate) to enhance the robustness, reliability, and clinical relevance of generated responses. By addressing these directions, we aim to further refine DermETAS-SNA into a comprehensive and reliable AI assistant for dermatological diagnosis and decision support in clinical use.

\begin{credits}
\subsection*{Acknowledgments}
The authors would like to thank Mr.Ermal Toto, Director, Scientific Data, Applications and Web Development  in the WPI Academic \& Research Computing group at Worcester Polytechnic Institute, for providing consulting support that contributed to the results presented in this work.
We extend our thanks to Dr. Sriram Vemula, Dr. Praveen R, Dr. Vishnukumar Subramanian, Dr. Thirumugilan B C, Dr. Saran Anand Murugan, Dr. Infant Nishanth A, and Dr. Raaga Likhitha Musunuri for their expert evaluation of clinical responses in this study.

\end{credits}
%
%
%

\begin{thebibliography}{99}

\bibitem{who2025skin} 
World Health Organization:
\textit{EB156 draft decision on skin diseases as a global public health priority}.
2025.
\url{https://cdn.who.int/media/docs/default-source/ntds/skin-ntds/global-meeting-on-skin-related-neglected-tropical-diseases-2025/eb156-draft-decision-on-skin-diseases-as-a--global-public-health-priority.pdf?sfvrsn=b64f45e8_3} (accessed July 19, 2025).


\bibitem{Tian2023} 
Tian, J., Zhang, D., Yang, Y., Huang, Y., Wang, L., Yao, X., Lu, Q.:
\textit{Global epidemiology of atopic dermatitis: a comprehensive systematic analysis and modelling study}.
British Journal of Dermatology \textbf{190}(1), 55--61 (2023).
\url{https://doi.org/10.1093/bjd/ljad339}

\bibitem{ref3} 
Maghiar, L.; Sandor, M.; Sachelarie, L.; Bodog, R.; Huniadi, A. Skin Lesions Caused by HPV—A Comprehensive Review. Biomedicines 2024, 12, 2098. \url{https://doi.org/10.3390/biomedicines12092098}


\bibitem{actinic} George CD, Lee T, Hollestein LM, Asgari MM, Nijsten T. Global epidemiology of actinic keratosis in the general population: a systematic review and meta-analysis. Br J Dermatol. 2024 Mar 15;190(4):465-476. doi: 10.1093/bjd/ljad371.

\bibitem{ref6} Greco MJ, Bhutta BS. Seborrheic Keratosis. [Updated 2024 May 6]. In: StatPearls [Internet]. Treasure Island (FL): StatPearls Publishing; 2025 Jan-. Available from: \url{https://www.ncbi.nlm.nih.gov/books/NBK545285/}

\bibitem{ref7} Bodman MA, Syed HA, Krishnamurthy K. Onychomycosis. [Updated 2024 Jan 9]. In: StatPearls [Internet]. Treasure Island (FL): StatPearls Publishing; 2025 Jan-. Available from: \url{https://www.ncbi.nlm.nih.gov/books/NBK441853/}

\bibitem{ref_article1} Siegel, R.L., Kratzer, T.B., Giaquinto, A.N., Sung, H., Jemal, A.: Cancer statistics, 2025. \textit{CA Cancer J Clin.} \textbf{75}(1), 10--45 (2025). \doi{10.3322/caac.21871}

\bibitem{Shah2024}
Shah M, Burshtein J, Zakria D, Rigel D. Analysis of trends in US dermatologist density and geographic distribution. \textit{J Am Acad Dermatol}. 2024 Aug;91(2):338--341. doi:10.1016/j.jaad.2024.03.037. PMID: 38574771.

\bibitem{skingpt2024}
Zhou, J., He, X., Sun, L., Xu, J., Chen, X., Chu, Y., Zhou, L., Liao, X., Zhang, B., Afvari, S., Gao, X.: Pre-trained multimodal large language model enhances dermatological diagnosis using SkinGPT‑4. \textit{Nature Communications} \textbf{15}(1), 5649 (2024). \doi{10.1038/s41467-024-50043-3}

\bibitem{vit}
A.~Dosovitskiy, L.~Beyer, A.~Kolesnikov,
``An Image is Worth 16x16 Words: Transformers for Image Recognition at Scale,''
\textit{arXiv preprint arXiv:2010.11929}, 2020.
Available: \url{https://doi.org/10.48550/arXiv.2010.11929}

\bibitem{llama2}
Hugo Touvron, Louis Martin, Kevin Stone, et.al
\textit{Llama 2: Open Foundation and Fine-Tuned Chat Models}, arXiv preprint arXiv:2307.09288, 2023. Available at: \url{https://doi.org/10.48550/arXiv.2307.09288}

\bibitem{skincon}
Daneshjou, R., Yuksekgonul, M., Cai, Z., Novoa, R.A., Zou, J.: SkinCon: A skin disease dataset densely annotated by domain experts for fine-grained model debugging and analysis. \textit{arXiv preprint} arXiv:2302.00785 (2023). \doi{10.48550/arXiv.2302.00785}

\bibitem{dermnet}
Goel, S.: DermNet Skin Disease Image Dataset. \textit{Kaggle} (2022). \url{https://www.kaggle.com/datasets/shubhamgoel27/dermnet}

\bibitem{rag}
Patrick Lewis, Ethan Perez, Aleksandra Piktus, et Al
\newblock Retrieval‑Augmented Generation for Knowledge‑Intensive NLP Tasks.
\newblock In *Advances in Neural Information Processing Systems (NeurIPS)*, vol.33, pp.9459–9474, 2020.  
\newblock arXiv:2005.11401. :contentReference[oaicite:1]{index=1}

\bibitem{gemini25pro}
Google Cloud: Gemini 2.5 Pro Preview 05–06. \url{https://cloud.google.com/vertex-ai/generative-ai/docs/models/gemini/2-5-pro}

\bibitem{habif2021}
Dinulos, J.G.H., and Habif, T.P. \textit{Habif’s Clinical Dermatology: A Color Guide to Diagnosis and Therapy}. 7th ed., Edinburgh: Elsevier, 2021.

\bibitem{rook2017}
Paul, C. \textit{Rook's Textbook of Dermatology}, 9th edn. Edited by Christopher Griffiths, Jonathan Barker, 2016; 4696 pp. ISBN: 978‐1118441190. \textit{British Journal of Dermatology}, 176(6), 1676--1677, 2017. \doi{10.1111/bjd.15624}.

\bibitem{primarycare2021}
Buckley, D., and Pasquali, P. (Eds.). \textit{Textbook of Primary Care Dermatology}. Cham: Springer, 2021.

\bibitem{fitzpatrick2008}
Wolff, K., Goldsmith, L., Katz, S., Gilchrest, B., Paller, A.S., and Leffell, D. \textit{Fitzpatrick's Dermatology in General Medicine}. 7th ed., McGraw-Hill, 2008.

\bibitem{electronic}
Drugge, R., Dunn, H. A., \& Internet Dermatology Society. \textit{The Electronic Textbook of Dermatology}. Internet Dermatology Society, 2000.

\bibitem{lama_parser}
LlamaIndex: LlamaParse – parse \& clean your data for RAG. \url{https://docs.llamaindex.ai/en/stable/llama_cloud/llama_parse/}

\bibitem{qwen}
Li, Z., Zhang, X., Zhang, Y., Long, D., Xie, P., Zhang, M.: Towards general text embeddings with multi-stage contrastive learning. \textit{arXiv preprint} arXiv:2308.03281 (2023)

\bibitem{qdrant}
Qdrant developers: Qdrant: Open-source Vector Database. Version 1.5.0 (2025). \url{https://qdrant.tech}

\bibitem{cohere}
Cohere: Introducing Rerank3.5: Precise AI Search. \url{https://cohere.com/blog/rerank-3pt5}

\bibitem{gpt4o}
OpenAI: GPT‑4o System Card. \url{https://openai.com/index/hello-gpt-4o/}

\bibitem{llama4}
Meta AI: LLaMA4Maverick – Model Card. \url{https://huggingface.co/meta-llama/Llama-4-Maverick-17B-128E-Instruct}

\bibitem{llmblender}
Dongfu Jiang, Xiang Ren, and Bill Yuchen Lin.
\newblock {LLM-Blender: Ensembling Large Language Models with Pairwise Ranking and Generative Fusion}.
\newblock \emph{arXiv preprint arXiv:2306.02561}, 2023.
\newblock URL: \url{https://doi.org/10.48550/arXiv.2306.02561}.

\bibitem{ENAS}
S. Deng, Z. Lv, E. Galván and Y. Sun, "Evolutionary Neural Architecture Search for Facial Expression Recognition," in IEEE Transactions on Emerging Topics in Computational Intelligence, vol. 7, no. 5, pp. 1405-1419, Oct. 2023, doi: 10.1109/TETCI.2023.3289974

\bibitem{resnet-50}
N. Gaffoor and S. Soomro, "Skin Disease Detection and Classification Using ResNet-50 and Support Vector Machine: An Effective Approach for Dermatological Diagnosis," 2023 IEEE International Conference on Internet of Things and Intelligence Systems (IoTaIS), Bali, Indonesia, 2023, pp. 140-145, doi: 10.1109/IoTaIS60147.2023.10346059.
\end{thebibliography}

\end{document}